\documentclass[final,3p,times]{elsarticle}
\def\SM{$\mathrm{SU(3)_c \otimes SU(2)_L \otimes U(1)_Y}$ }

\def\lfv{lepton flavour violation }
\def\lnv{lepton number violation }

\usepackage{amssymb}

\newcommand{\AddrAHEP}{AHEP Group, Institut de F\'{i}sica Corpuscular --
  C.S.I.C./Universitat de Val\`{e}ncia, Parc Cientific de Paterna.\\
  C/Catedratico Jos\'e Beltr\'an, 2 E-46980 Paterna (Val\`{e}ncia) - SPAIN}

\newcommand{\Cinvestav}{Departamento de F\'{\i}sica, Centro de
  Investigaci{\'o}n y de Estudios Avanzados del IPN\\ Apdo. Postal
  14-740 07000 Mexico, Distrito Federal, Mexico}

\def\vev#1{\left\langle #1\right\rangle}
\def\SM{$\mathrm{SU(3)_c \otimes SU(2)_L \otimes U(1)_Y}$ }
\def\21{$\mathrm{SU(2)_L \otimes U(1)_Y}$ }
\def\lfv{lepton flavour violation }
\def\lnv{lepton number violation }

\def\TrTrOne{ $\mathrm{SU(3)_c \otimes SU(3)_L \otimes U(1)_X}$ }
\def\lfv{lepton flavour violation }
\def\lnv{lepton number violation }
\def \znbb {$\rm 0\nu\beta\beta$ }
\newcommand{\sm}{standard model }
\newcommand {\ignore}[1]{}

\usepackage{color}
\usepackage[normalem]{ulem}

\journal{Nuclear Physics B}

\begin{document}

\begin{frontmatter}



\title{Neutrino oscillations  and the seesaw origin of neutrino mass}


\author{O. G. Miranda}
\address{\Cinvestav}
\author{J. W. F. Valle}
\address{\AddrAHEP}

\begin{abstract}

  The historical discovery of neutrino oscillations using solar and
  atmospheric neutrinos, and subsequent accelerator and reactor
  studies, has brought neutrino physics to the precision era.
  We note that CP effects in oscillation phenomena could be difficult
  to extract in the presence of unitarity violation.
  As a result upcoming dedicated leptonic CP violation studies should
  take into account the non-unitarity of the lepton mixing matrix.
  Restricting non-unitarity  will shed light on the seesaw
  scale, and thereby guide us towards the new physics responsible for
  neutrino mass generation.
\end{abstract}

\begin{keyword}
Neutrino mass \sep Neutrino mixing \sep Neutrino interactions



\end{keyword}

\end{frontmatter}

\section{Introduction}
\label{Introduction}

Particle physics has seen two historic discoveries in less than twenty
years: the confirmation of the mechanism of electroweak symmetry
breaking~\cite{Aad:2012tfa} and the discovery of neutrino
oscillations~\cite{fukuda:1998mi,ashie:2004mr,fukuda:2002pe,ahmad:2002jz,eguchi:2002dm},
both deservedly honored with the Nobel prize.
The unification paradigm~\cite{Georgi:1974sy,PhysRevLett.33.451} and
the good behaviour of the electroweak breaking sector, including
naturalness, perturbativity and stability~\cite{Djouadi:2005gi}, have
so far provided a strong theoretical motivation for new physics.
Other hints are the understanding of flavour and the unification with
gravity, in addition to the challenges posed by cosmological
observations associated to dark matter, dark energy and inflation.
Last, but not least, the need to account for non-zero neutrino
mass~\cite{Maltoni:2004ei,Forero:2014bxa} plays a key role in
the quest for new physics~\cite{Valle:2015pba}.

The most popular mechanism of neutrino mass generation ascribes the
smallness of neutrino mass as resulting from the exchange of heavy
messenger particles, such as right-handed iso-singlet neutrinos and/or
iso-triplet scalar bosons, known as the seesaw
mechanism~\cite{Valle:2015pba}. When formulated at low-scale this
naturally implies new effects in neutrino propagation that go beyond
the oscillatory behaviour, as explained below. In particular, future
neutrino experiments will face the challenge of disentangling
``conventional'' CP violation with that associated to the
non-unitarity of the lepton mixing matrix, which in turn results as an
indirect effect of the extra neutral heavy right-handed neutrinos.

In what follows, we briefly review current neutrino oscillation
parameters and describe novel effects associated to right-handed
neutrino admixture in the charged current weak interaction, expected
in low-scale seesaw schemes, purely in the context of the \SM
paradigm. We also recompile current limits on right-handed neutrino
mass and mixing parameters. 

\section{Three neutrino mixing and oscillations}
\label{sec:struct-lept-mixing}

Generic neutrino mass schemes require interactions associated to new
Yukawa couplings that do not commute with those of the charged
leptons, leading to the phenomenon of mixing in the charged current
weak interactions, analogous to the CKM mixing of
quarks~\cite{kobayashi:1973fv}. However, as we will see its structure
can be richer.

\subsection{Lepton mixing matrix for Dirac neutrinos}
\label{sec:lepton-mixing-matrix-d}

The mixing of leptons arising from the non-simultaneous
diagonalizability of the Dirac neutrino and charged lepton mass
matrices is given by an arbitrary unitary matrix
\begin{equation}
U = \omega_0(\gamma)  \prod_{i<j}^{3} \omega_{ij}(\eta_{ij})~,
\end{equation}
 where each of the $\omega$ factors is effectively $2\times 2$,
  characterized by an angle and a corresponding CP phase, e.g.
\begin{equation}
   \label{eq:w13}
\omega_{13} = \left(\begin{array}{ccccc}
c_{13} & 0 & e^{-i \phi_{13}} s_{13} \\
0 & 1 & 0 \\
-e^{i \phi_{13}} s_{13} & 0 & c_{13}
\end{array}\right)~,  \nonumber
 \end{equation}
 while $\omega_0 (\gamma )$ is an arbitrary diagonal unitary matrix.
 In complete analogy with the \sm quark sector we can use the phase
 redefinition freedom for neutral and charged leptons to show that
 only one independent CP~phase remains for the three Dirac
 neutrinos. To find this rephasing invariant parameter we use the
 conjugation property~\cite{Schechter:1980gr}
\begin{equation}
P^{-1} \, U \, P = 
 \omega_{23}(\theta_{23}; \phi_{23} - \beta)~
\omega_{13}(\theta_{13}; \phi_{13} - \alpha)~\omega_{12}(\theta_{12};
\phi_{12} + \beta -\alpha) \,,
  \end{equation}
  with $P = diag(e^{i\alpha},e^{i\beta},1)$, which allows us to
  identify~\cite{Rodejohann:2011vc},
\begin{equation} \label{eq:delta}
\delta \leftrightarrow \phi_{13} -\phi_{12} - \phi_{23} \, .
\end{equation}
The form of the lepton mixing matrix in this case is the same as the
CKM matrix describing the mixing of quarks.

\subsection{Lepton mixing matrix for Majorana neutrinos}
\label{sec:lepton-mixing-matrix-m}

The case of 3$\times$3 unitary lepton mixing matrix arises, for
example, if the light neutrino masses result from the exchange of
iso-triplet scalar messengers through Type II seesaw
~\cite{Schechter:1980gr} (see below).
Within the symmetric parametrization the mixing matrix has the form
\begin{equation} \label{eq_jv}
U = \omega_{23}(\theta_{23}; \phi_{23})~\omega_{13}(\theta_{13}; \phi_{13})~\omega_{12}(\theta_{12}; \phi_{12}) \, , 
  \end{equation}
  where the diagonal unitary matrix $\omega_0 (\gamma )$ is eliminated
  by rephasing the charged leptons but no rephasing of the neutrinos
  is possible, leading to two extra phases in the mixing matrix $U$
  compared with the previous case~\cite{Schechter:1980gr}. Explicitly,
  the matrix $U$ can be written as:
\begin{equation}
U=\left( \begin{array}{c c c}
c_{12}c_{13}&s_{12}c_{13}e^{-i{\phi_{12}}}&s_{13}e^{-i{\phi_{13}}}\\
-s_{12}c_{23}e^{i{\phi_{12}}}-c_{12}s_{13}s_{23}e^{-i({\phi_{23}}-{\phi_{13}})}
&c_{12}c_{23}-s_{12}s_{13}s_{23}
e^{-i({\phi_{12}}+{\phi_{23}}-{\phi_{13}})}&c_{13}s_{23}e^{-i{\phi_{23}}}\\
s_{12}s_{23}e^{i({\phi_{12}}+{\phi_{23}})}-c_{12}s_{13}c_{23}e^{i{\phi_{13}}}
&-c_{12}s_{23}e^{i{\phi_{23}}}-
s_{12}s_{13}c_{23}e^{-i({\phi_{12}}-{\phi_{13}})}&c_{13}c_{23}\\
\end{array} \right).
\label{eq:writeout}
\end{equation}
Although they do not show up in oscillations, the ``Majorana'' phases
will affect \lnv processes such as
\znbb~\cite{Schechter:1980gk,Doi:1980yb}. It has been noticed that
this fully symmetrical presentation is more convenient for the
description of \znbb decay that the PDG form~\cite{Rodejohann:2011vc}
(see below).

\subsection{Status of the three neutrino picture}
\label{sec:stat-neutr-oscill}

Neutrinos from natural sources like the Sun, and from the interaction
of cosmic rays with the Earth's atmosphere gave us the first
indications for neutrino conversion.
Neutrinos are also produced in the laboratory, both at accelerators as
well as nuclear reactors. 
The disappearance of muon neutrinos over a long-baseline probing the
same region of squared mass splitting relevant for atmospheric
neutrinos has been obtained in accelerator neutrino oscillation
experiments, starting with the KEK to Kamioka (K2K) neutrino oscillation
experiment, the MINOS Experiment using the NuMI Beam-line facility at
Fermilab, and currently the T2K (Tokai to Kamioka) experiment in Japan
and the NOvA experiment in the USA.
These have also substantially helped in determining the neutrino
parameters with a high level of
precision~\cite{Abe:2011sj,Abe:2013hdq,Adamson:2011qu,Adamson:2013whj,Adamson:2013ue}.

The results of solar neutrino experiments have also been confirmed by
reactor neutrino studies at the KamLAND experiment~\cite{eguchi:2002dm}.
More sensitive experiments such as Double Chooz~\cite{Abe:2011fz},
RENO~\cite{Ahn:2012nd}, and Daya Bay~\cite{An:2012eh} have confirmed
that $\theta_{13}$ is nonzero. Particularly Daya Bay has now provided a
precision measurement of $\theta_{13}$~\cite{An:2012eh}, one of the
most important results in the field in this decade.
\begin{figure}[htb] \centering
   \includegraphics[width=.5\linewidth]{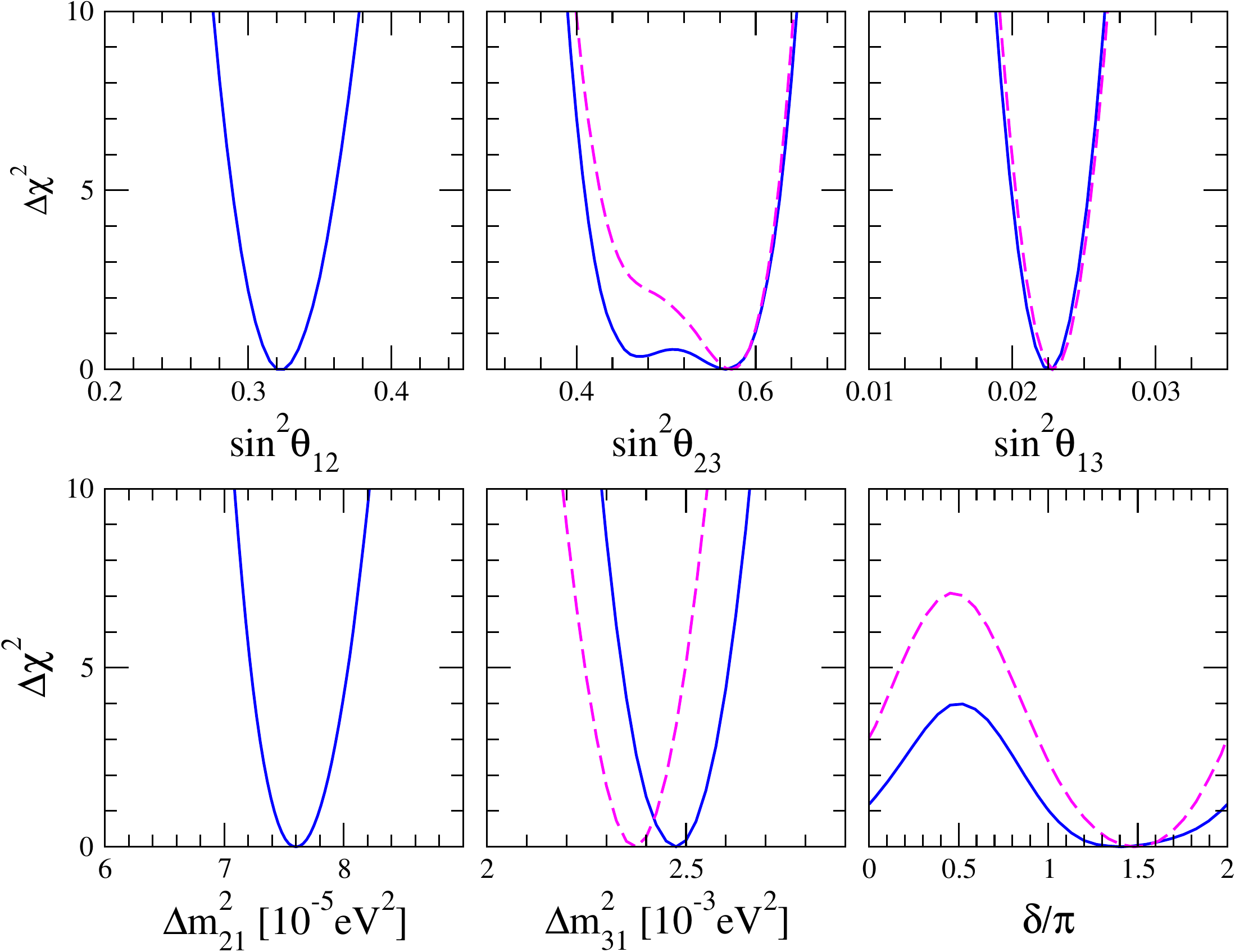}
   \caption{\label{fig:status} Neutrino mixing angles and squared mass
     differences, as determined from the global analysis in
     Ref.~\cite{Forero:2014bxa}.  The normal hierarchy case is shown
     with solid lines; dashed lines are used for inverted hierarchy.}
\end{figure}
Altogether, neutrino physics is now a mature branch of science, in
which the three neutrino mixing angles as well as the two squared mass
differences have been determined with high precision.
The current status of the determination of the solar, atmospheric,
reactor and accelerator neutrino oscillation parameters is summarized
in Fig.~(\ref{fig:status}), taken from Ref.~\cite{Forero:2014bxa},
where the solid and dashed curves refer to the normal and inverted
mass ordering.
These include data from a number of solar neutrino
experiments~\cite{Aharmim:2009gd,aharmim:2008kc,Abe:2010hy,Renshaw:2014awa,Hosaka:2005um,Cravens:2008aa,Bellini:2011rx,Abdurashitov:2009tn,Kaether:2010ag,cleveland:1998nv}
and atmospheric data from the Super-Kamiokande
experiment, described in Ref.~\cite{Wendell:2010md}.
All of these experiments have been taken into account in the results
summarized here, as well as in other similar studies, all of which
agree at the $3\sigma$
level~\cite{Capozzi:2013csa,Gonzalez-Garcia:2014bfa}.
One sees that for the case of $\theta_{23}$ there is still room for
improvement. Concerning the standard Dirac CP phase, $\delta$ in
Eq.~(\ref{eq:delta}), at the moment there is only a hint that it is
non-vanishing.
Finally, notice that oscillation experiments provide no information on
the absolute neutrino mass scale, nor on the values of the Majorana
phases. 

\subsection{Robustness of neutrino oscillations}
\label{sec:robustn-w.r.t.-unit}

How robust is the oscillation interpretation of current neutrino data?
So profound is the discovery of neutrino oscillations and the
determination of neutrino oscillation parameters that it requires
careful consideration of any possible loopholes. 
The good agreement between the standard solar model sound speed
profile and helioseismology results substantially constrain possible
astrophysical
uncertainties~\cite{Robertson:2012ib,Balantekin:2013tqa}.
Yet the effect of varying solar neutrino fluxes has been widely
discussed, without substantial impact on the neutrino
oscillation parameters.
However, although experiments are now measuring neutrino fluxes to
within a few percent, helioseismic studies have reached accuracies of
about a few parts in a thousand. Hence, it is not inconceivable that
discrepancies might eventually show
up~\cite{Robertson:2012ib,Balantekin:2013tqa}.

Uncertainties associated with the possibility of solar density
fluctuations were first suggested in the late
nineties~\cite{balantekin:1996pp,nunokawa:1996qu}.
Such fluctuations deep within the Sun could have a resonant origin
from magnetic fields in the radiative zone~\cite{burgess:2003fj}.
Direct helioseismic tests are not necessarily in conflict with such
variations, since they are not sensitive to fluctuations with size
around several hundreds of kilometers to which neutrino oscillations
are sensitive~\cite{Castellani:1997pk,Christensen-Dalsgaard:2002ur}.
Indeed, it has been shown that the effect on solar neutrino
oscillations can be important. However, the measurement of neutrino
properties at KamLAND provides valuable independent information which
can, in fact, be used to probe the deep solar
interior~\cite{burgess:2002we}.  Solar neutrino measurements from SNO
are now sufficiently precise that neutrino oscillation parameters can
be inferred independently of any assumptions about fluctuation
size~\cite{burgess:2003su}. In fact the fluctuation amplitudes above
5\% now excluded if their correlation lengths lie in the range of
several hundred km.

Magnetic fields in the solar convective zone can cause spin-flavour
precession and produce a solar $\bar\nu_e$
flux~\cite{miranda:2000bi,barranco:2002te}. The robustness of the
oscillation hypothesis has also been analysed in this context. It has
been shown that $\bar\nu_e$ production can be greatly enhanced for the
case of random magnetic fields~\cite{miranda:2004nz}.
The search for anti-neutrinos from the Sun can be used to constrain
the neutrino magnetic moments~\cite{miranda:2003yh}.
In summary, laboratory oscillation studies not only give a crucial
confirmation of the solar neutrino oscillation hypothesis, ruling out
exotic solutions, but also establish the robustness of large mixing
angle oscillations~\cite{Maltoni:2002aw,pakvasa:2003zv}.

\section{Seesaw paradigm}
\label{sec:seesaw-schemes}

As the only electrically neutral \sm fermions, neutrino mass
generation could be different from the standard Higgs mechanism. In
particular neutrinos could be light as a result of their Majorana
nature.
Indeed, unless prevented by basic symmetries such as electric or
colour charge~\cite{Schechter:1980gr}, fermions are intrinsically
two-component objects. The emergence of Dirac neutrinos, would then
signal extra symmetry assumptions. For example one could assume lepton
number conservation directly, or some extended symmetry that implies
the conservation of lepton number, as some specific flavour
symmetries~\cite{Aranda:2013gga}. Moreover, following Weinberg, we
note the simplest operator capable of inducing neutrino masses is a
unique dimension-5 operator, generally implying \lnv and Majorana
neutrinos. In fact mechanisms with Dirac neutrino masses might be an
indication for physics beyond four dimensions~\cite{Chen:2015jta}.

\begin{figure}[htp] \centering
   \includegraphics[width=.45\linewidth]{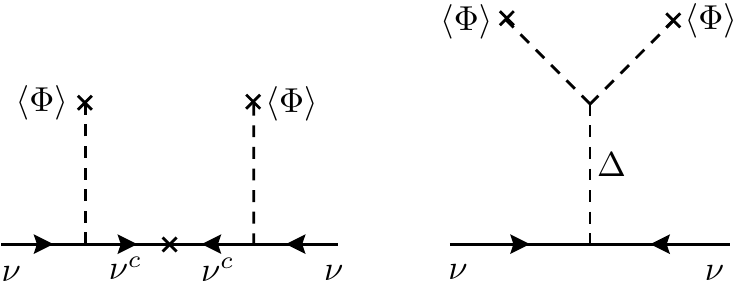}
   \caption{\label{fig:seesaw} In the Type-I seesaw the neutrino mass
     is induced from the fermion exchange, while Type-II corresponds
     to scalar boson exchange.  }
\end{figure}

The most popular way to induce Weinberg's dimension-5 operator is
through the so--called seesaw mechanism, which represents a huge
variety of possible schemes. The first case is called pure Type-I
while the second is pure Type-II, a terminology opposite of the
original one suggested in~\cite{Schechter:1980gr}, see
Fig.~\ref{fig:seesaw}.
Note that the Type-I seesaw mechanism corresponds to having the
neutrino mass induced from fermion exchange. In this case, as we will
see, neutrino oscillations are effectively described by a non-unitary
lepton mixing matrix.

Given the arbitrariness in the number and transformation properties of
extra fermions (for example, the number of right-handed neutrinos is
arbitrary, since they carry no anomaly~\cite{Schechter:1980gr}) the
seesaw mechanism can be realized at low
scale~\cite{Mohapatra:1986bd,gonzalezgarcia:1989rw,Akhmedov:1995ip,Akhmedov:1995vm,Malinsky:2005bi}.
In this case the messenger particles may be indirectly probed through
rare \lfv decay
processes~\cite{bernabeu:1987gr,Deppisch:2013cya,Deppisch:2004fa,Deppisch:2005zm,Abada:2014kba,Boucenna:2015zwa}
and electroweak precision physics~\cite{PhysRevLett.65.964,
  ALTARELLI1991161, PhysRevD.46.381}, or be directly produced at
collider experiments~\cite{Das:2012ii,Deppisch:2015qwa}.

Before closing this section, let us mention that seesaw extensions of
the \sm may have deep implications for new physics, ranging from
neutrino physics \textit{stricto sensu} to other aspects of particle
physics and cosmology.
In particular, the presence of new scalars required in order to break
lepton number may affect the stability of the electroweak
breaking~\cite{Bonilla:2015kna,Bonilla:2015eha}.
Moreover, extra scalars can also induce new contributions to
``visible'' \sm Higgs decays, such as the $h\to\gamma\gamma$ and
possibly account for new hints, such as the recent diphoton
anomaly~\cite{Boucenna:2015pav,Dong:2015dxw}. There may also be novel
Higgs decay channels involving the emission of the Nambu-Goldstone
boson associated to spontaneous \lnv and neutrino mass
generation~\cite{joshipura:1992hp,Diaz:1998zg,Bonilla:2015uwa}.

\section{Non-unitary lepton mixing and seesaw mechanism}
\label{sec:lepton-mixing-seesaw}

The two-component right-handed neutrinos are singlets under the \SM
symmetry and hence can acquire potentially large gauge invariant
masses, breaking total lepton number symmetry. This opens the
possibility of generating light neutrino Majorana masses through the
exchange of heavy right-handed neutrinos, the so-called Type-I seesaw
mechanism. This implies that, in addition to the presence of Majorana
phases, the lepton mixing matrix will also couple sub-dominantlly the
heavy states, leading to a rectangular form for the ``PMNS''
matrix~\cite{Schechter:1980gr}.  These couplings enable the production
of the right-handed neutrinos by the charged current weak interaction
if the kinematics allows, possibly at the LHC.
In most other cases, however, such as oscillations, the heavy states
can not participate due to kinematics.
As a result, the charged current weak interactions of the light
(mass-eigenstate) neutrinos are described by
an effective non-unitary mixing matrix.\\[-.3cm]

In order to find the most convenient parametrization for the matrix
$N$ describing non-unitary neutrino propagation we start from the
unitary matrix $U^{n\times n}$ describing the neutrino diagonalization
matrix. In the symmetric parametrization the products of the complex
matrices $\omega_{ij}$ in Eq.~(\ref{eq:w13}) can be chosen in the
most convenient way as
\begin{equation}
U^{n\times n}=\omega_{n-1\, n}\:\omega_{n-2\, n}\:\ldots\:\omega_{1\, n}\:\omega_{n-2\, n-1}\:\omega_{n-3\, n-1}\:\ldots\:\omega_{1\, n-1}\:\ldots\:\omega_{2\,3}\:\omega_{1\,3}\:\omega_{1\,2}\label{eq:Ueq} \, .
\end{equation}
Following the notation in~\cite{Hettmansperger:2011bt} we break up
this matrix $U^{n\times n}$ as
\begin{equation}
U^{n\times n}=\left(\begin{array}{cc} N & S\\
V & T
\end{array}\right)\label{eq:ULindner_C1} ,
\end{equation}
where $N$ is a $3\times3$ matrix in the light neutrino sector and $S$
describes the couplings of the extra $n-3$ isosinglet states, expected
to be heavy. The matrix $U^{n\times n}$ can be neatly expanded in
perturbation theory~\cite{Schechter:1981cv}.
With the ordering shown in Eq.~(\ref{eq:Ueq}) the submatrix $N$ can be
decomposed as
\begin{equation}
N=N^{NP}\, U=\left(\begin{array}{ccc}\alpha_{11} & 0 & 0\\
\alpha_{21} & \alpha_{22} & 0\\
\alpha_{31} & \alpha_{32} & \alpha_{33}
\end{array}\right)\: U
\label{eq:Ndescopm_C1} ,
\end{equation}
with $U$ the usual unitary $3\times 3$ leptonic mixing matrix. The
latter may be expressed as prescribed by the Particle Data
Group~\cite{Agashe:2014kda} or in our fully symmetric description,
particularly useful for our analyses~\cite{Rodejohann:2011vc}, namely,
\begin{equation}
U =  \omega_{2\,3}\:\omega_{1\,3}\:\omega_{1\,2}.
\end{equation}
Note that Eq.~(\ref{eq:Ndescopm_C1}) gives the most general and
convenient description of the lepton mixing matrix relaxing the
unitarity approximation, and holds in any seesaw scheme.
Notice that in this factorized form the left pre-factor matrix,
$N^{NP}$, characterizes the unitarity violation.  Notice that the
oscillations of the electron and muon neutrino flavors are described
by just four extra parameters, the two real diagonal entries
$\alpha_{11}$ and $\alpha_{22}$ plus the complex off-diagonal
parameter $\alpha_{21}$ which contains a single additional effective
CP phase parameter. In other words only one phase combination enters
the ``relevant'' neutrino oscillation experiments (next section). This
important property holds irrespective of the number of extra heavy
leptons present.
Hence, by conveniently choosing the product ordering of the complex
rotation matrices, we obtained a parametrization that concentrates all
the information relative to the additional neutral heavy leptons in a
compact and simple matrix that contains three zeroes.  \\[-.3cm]

Notice also that, because of the chiral nature of the \SM model, the
couplings of the $n$ neutrino states in the charged current weak
interaction will form a rectangular matrix~\cite{Schechter:1980gr},
that is,
\begin{equation}
K=\left(\begin{array}{cc} N & S
\end{array}\right)\label{eq:K} . 
\end{equation}
where the first block corresponds to the three active neutrinos and
the second is associated to the other states. The unitarity condition
will be replaced by the relation
\begin{equation}
KK^\dagger = NN^\dagger + SS^\dagger = I ,
\label{eq:unita}
\end{equation}
with 
\begin{equation}
NN^\dagger=\left(\begin{array}{lcccl} \alpha_{11}^2\ & & 
                                 \alpha_{11}\alpha^*_{21} &  &
                                 \alpha_{11}\alpha^*_{31} \\
      \alpha_{11}\alpha_{21} & &
                    \alpha_{22}^2  + |\alpha_{21}|^2 & &
                    \alpha_{22}\alpha^*_{32} + \alpha_{21}\alpha^*_{31}  \\
      \alpha_{11}\alpha_{31} & &
                    \alpha_{22}\alpha_{32} + \alpha_{31}\alpha^*_{21} & &
     \alpha_{33}^2 + |\alpha_{31}|^2 + |\alpha_{32}|^2 
\end{array}\right) \, . \label{eq:NN}
\end{equation}

Besides being described by a triangular matrix, another advantage of
the parametrization is that the large number of mixing angles and
phases coming from any number of extra heavy isosinglets can be
reconstructed by relatively simple formulas. This is particularly true
for the diagonal elements, $\alpha_{ii}$, which are expressed as
\begin{eqnarray}
\alpha_{11} \: &=& \: c_{1\, n}\: c_{\,1n-1}c_{1\, n-2}\ldots c_{14}  , \nonumber \\
\alpha_{22} \: &=& \: c_{2\, n}\: c_{\,2n-1}c_{2\, n-2}\ldots c_{24} , \\
\alpha_{33} \: &=& \: c_{3\, n}\: c_{\,3n-1}c_{3\, n-2}\ldots c_{34}  \nonumber ,
\end{eqnarray}
where ${c}_{ij}=\cos\theta_{ij}$, the cosines of the mixing
parameters are real.
For the off-diagonal terms, $\alpha_{21}$ and $\alpha_{32}$, there is
also a general and simple formula, given as the sum of $n-3$ terms
\begin{eqnarray}
\alpha_{21} \: = \:
    c_{2\, n}\: c_{\,2n-1}\ldots c_{2\, 5}\:{\eta}_{24}\bar{\eta}_{14}\: +\: 
    c_{2\, n}\: \ldots c_{2\, 6}\:{\eta}_{25}\bar{\eta}_{15}\:c_{14} +\: 
    \ldots\:+ {\eta}_{2n}\bar{\eta}_{1n}\:c_{1n-1}\:c_{1n-2}\:\ldots\:c_{14} 
\nonumber \, , \\
\alpha_{32} \: = \:
    c_{3\, n}\: c_{\,3n-1}\ldots c_{3\, 5}\:{\eta}_{34}\bar{\eta}_{24}\: +\: 
    c_{3\, n}\: \ldots c_{3\, 6}\:{\eta}_{35}\bar{\eta}_{25}\:c_{24} +\: 
    \ldots\:+ {\eta}_{3n}\bar{\eta}_{2n}\:c_{2n-1}\:c_{2n-2}\:\ldots\:c_{24} . 
\label{eq:alfa_crossed_C1}
\end{eqnarray}
These parameters will be complex, and the CP phase information will be
encoded in the parameters
${\eta}_{ij}=e^{-i\phi_{ij}}\,\sin\theta_{ij}$ and its conjugate
$\bar{\eta}_{ij}=-e^{i\phi_{ij}}\,\sin\theta_{ij}$.  Finally, for the
$\alpha_{31}$ case, we can neglect quartic terms in $\sin\theta_{ij}$,
with $j=4,5,\cdots$ and find the expression
\begin{equation}
\alpha_{31} \: = \:
    c_{3\, n}\: c_{\,3n-1}\ldots c_{3\, 5}\:{\eta}_{34} c_{2\, 4} \bar{\eta}_{14}\: +\: 
    c_{3\, n}\: \ldots c_{3\, 6}\:{\eta}_{35} c_{2\, 5}\bar{\eta}_{15}\:c_{14} +\: 
    \ldots\: 
    + \: {\eta}_{3n} c_{2\,
  n}\bar{\eta}_{1n}\:c_{1n-1}\:c_{1n-2}\:\ldots\:c_{14}  \, .
\label{eq:alfa_31}
\end{equation}

In order to generate a realistic neutrino spectrum at the tree level
the Type-I seesaw mechanism requires two right-handed
neutrinos. However in the presence of extra Higgs bosons (such as
triplets present in Type II seesaw) this is not required. For the
simple case of a single extra right-handed neutrino the above
expressions give
\begin{eqnarray}
 \alpha_{11} = c_{14}\nonumber \, 
,  &\,& \, \, \, \, 
  \alpha_{21} = {\eta}_{24}\,\bar{\eta}_{14} \nonumber\label{eq:alpha31} \, ,\\
 \alpha_{22} = c_{24} \, , &\,& \, \, \, \, 
 \alpha_{32} ={\eta}_{34}\,\bar{\eta}_{24} \, ,\\
 \alpha_{33} = c_{34}\nonumber \, ,&\,& \, \, \, \, 
 \alpha_{31} = {\eta}_{34}\, c_{24}\, \bar{\eta}_{14}\nonumber\, .
\end{eqnarray}
Specific expressions for various other interesting seesaw schemes with
$3$ and $6$ heavy leptons are also given in
Ref.~\cite{Escrihuela:2015wra}.
The couplings of the heavy right-handed neutrinos allow them to be
searched for in many experiments. Moreover their presence implies the
effective non-unitarity of the light neutrino mixing matrix, modifying
several \sm observables (see Sec.~\ref{sec:curr-constr-right}) as well
as oscillation probabilities which have a very simple form in vacuo as
seen in Sec.~\ref{sec:prob-non-unit}.

\section{Current constraints on right-handed neutrinos}
\label{sec:curr-constr-right}

Even though right-handed neutrinos as messengers of neutrino mass
generation, as postulated in seesaw type schemes, are expected to be
heavy, above the weak scale or so, they have been searched for in a
variety of situations, starting from much lower masses.  Here we start
by briefly update the constraints on right-handed neutrinos.

\subsection{Low-mass searches}
\label{sec:searches}

\begin{figure}[!h]
\centerline{
\includegraphics[width=.65\linewidth]{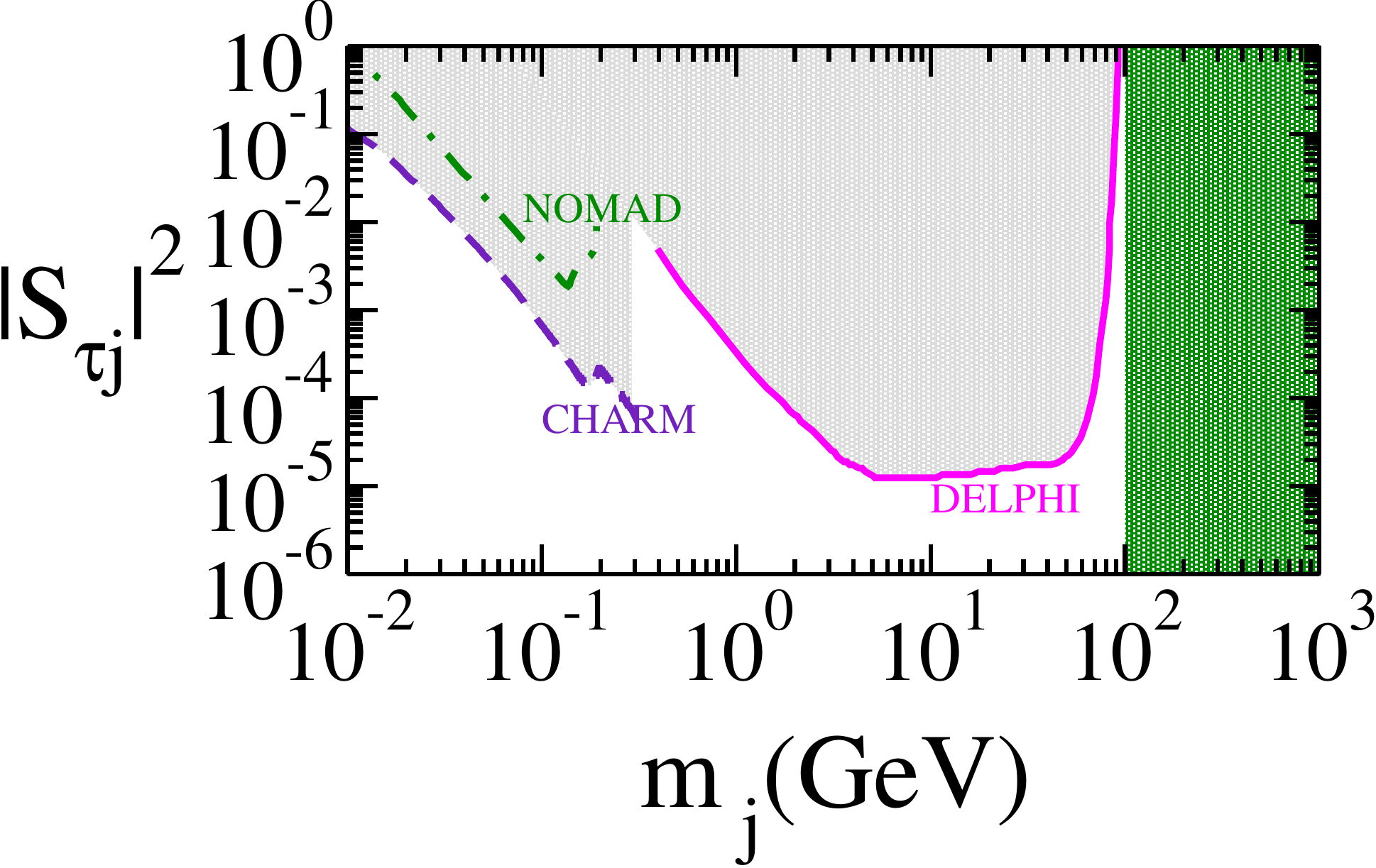}
}
\caption{\label{fig:Ktauj} Bounds on the charged current coupling
  strength of a heavy isosinglet lepton of mass $m_j$ with the tau
  neutrino. The green vertical band shows the lower part of the mass
  region favoured on theoretical grounds.}
   \end{figure}

If their mass is low enough, the right-handed states would behave as
light sterile neutrinos and would show up at low energies. Indeed, they
have been searched for in a variety of weak processes such as pion and
kaon weak decay as well as at the LEP
experiments~\cite{Dittmar:1989yg,Akrawy:1990zq}.
\begin{figure}[!h]
\centerline{
\includegraphics[width=.65\linewidth]{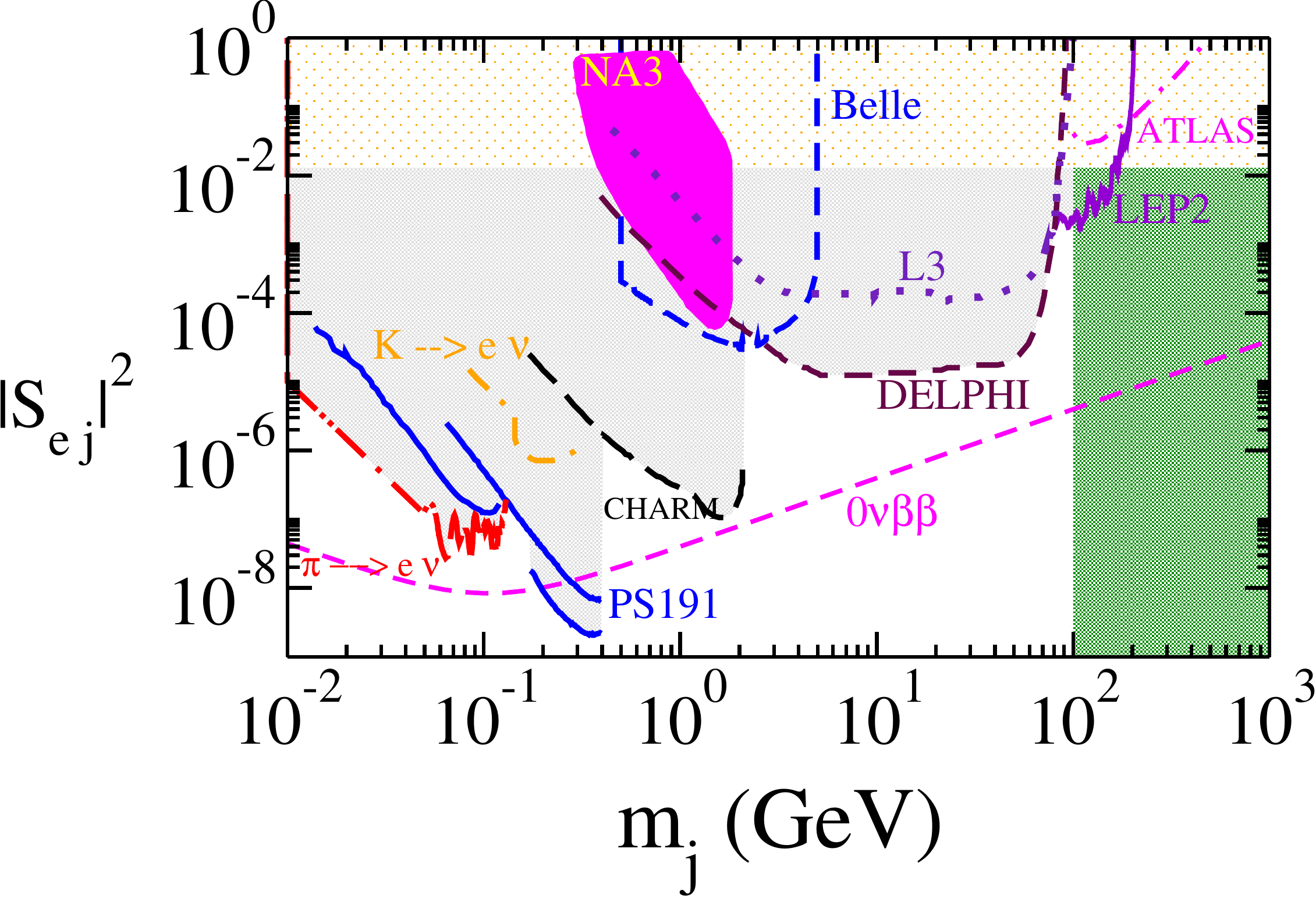}
}
\caption{\label{fig:Kej} Bounds on the charged current coupling
  strength of a heavy isosinglet lepton of mass $m_j$ with the
  electron neutrino. The green vertical band indicates the lower part
  of the mass region favoured on theoretical grounds, while the
  horizontal band shows the bound from weak universality.}
\end{figure}
\begin{figure}[!h]
\centerline{
\includegraphics[width=.65\linewidth]{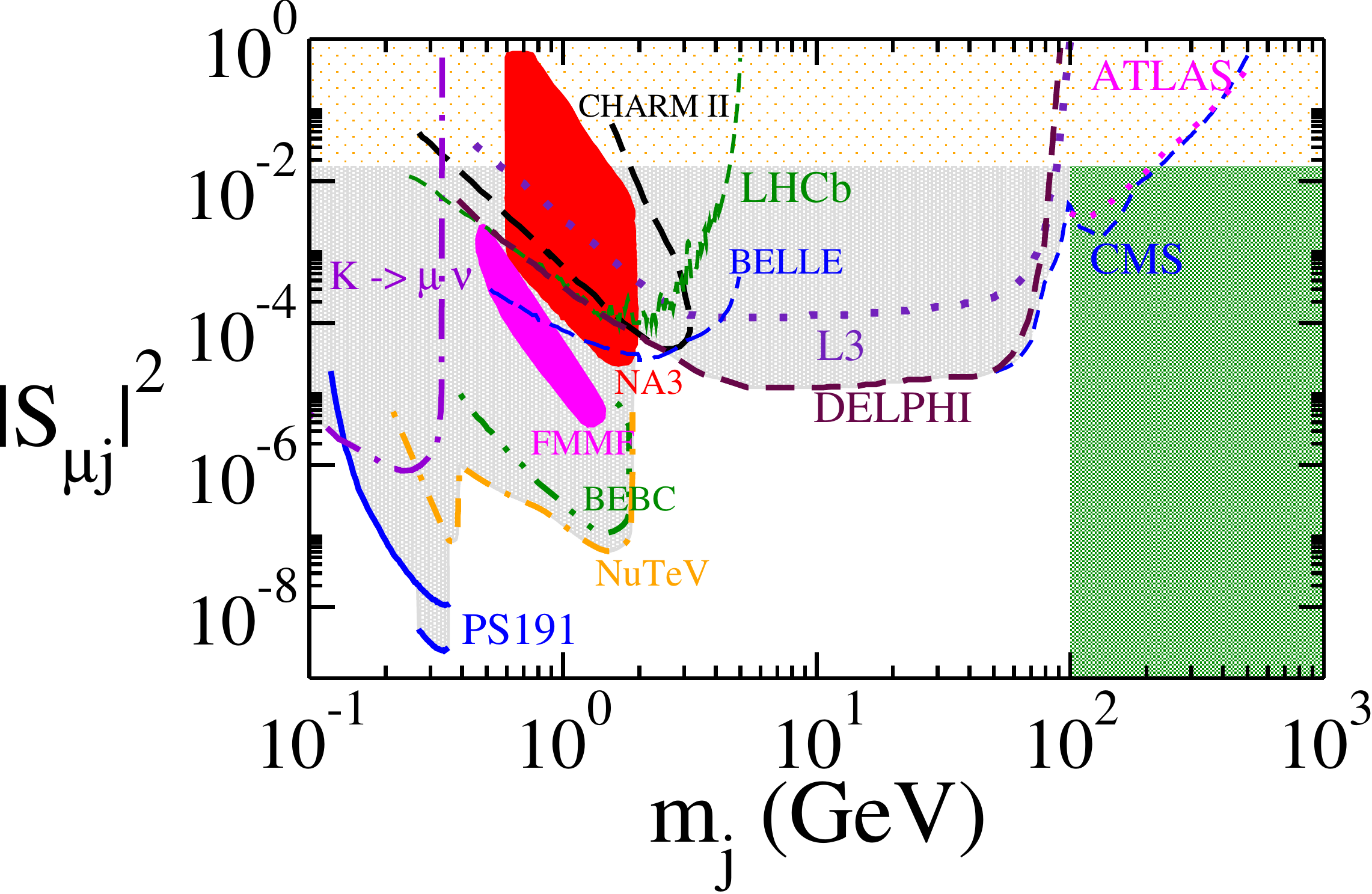}
}
\caption{\label{fig:Kmuj} Bounds on the charged current coupling
  strength of a heavy isosinglet lepton of mass $m_j$ with the muon
  neutrino.  The green vertical band indicates the lower part of the
  mass region favoured on theoretical grounds, while the horizontal
  band shows the bound from weak universality.}
\end{figure}
The constraints from direct production of neutral heavy leptons are
summarized in Fig.~(\ref{fig:Kej}), Fig.~(\ref{fig:Kmuj}) and
Fig.~(\ref{fig:Ktauj}).  These model-independent limits do not require
any particular neutrino mass generation scheme.  To obtain these
limits, experiments have looked for a resonance in a specific energy
window, for a given mixing of the additional heavy state.
In Fig.~(\ref{fig:Kej}), we summarize the constraints on $|S_{ej}|^2$
for a mass range from $10^{-2}$ to $10^2$~GeV coming from the
experiments TRIUMF~\cite{Britton:1992pg,Britton:1992xv} (denoted as
$\pi\to e\nu$ and $K\to e\nu$ in the plot),
PS191~\cite{Bernardi:1987ek}, NA3~\cite{Badier:1986xz},
CHARM~\cite{Bergsma:1985is}, Belle~\cite{Liventsev:2013zz}, the LEP
experiments DELPHI~\cite{Abreu:1996pa}, L3~\cite{Adriani:1992pq},
LEP2~\cite{Achard:2001qv}, and the recent LHC results from
ATLAS~\cite{ATLAS:2012yoa,klinger}.
Future experimental proposals, such as
DUNE~\cite{Adams:2013qkq,Acciarri:2015uup} and ILC, expect to improve
these constraints~\cite{Deppisch:2015qwa}.

In Fig.~(\ref{fig:Kmuj}) we show the constraints for the mixing of a
neutral heavy lepton with a muon neutrino.  The bounds from
experiments PS191, NA3, Belle, the LEP experiments L3, DELPHI, and the
LHC experiment ATLAS, are shown again for this channel. For the
muon-type neutrino, there are also bounds from
KEK~\cite{Hayano:1982wu,Kusenko:2004qc} (labeled $K\to \mu\nu$ in the
plot), CHARM II~\cite{Vilain:1994vg}, FMMF~\cite{Gallas:1994xp},
BEBC~\cite{CooperSarkar:1985nh}, NuTeV~\cite{Vaitaitis:1999wq},
E949~\cite{Artamonov:2014urb}, and the recent constraints from
the LHC experiments CMS~\cite{Khachatryan:2015gha} and
LHCb~\cite{Aaij:2014aba}.
Finally, we show in Fig.~(\ref{fig:Ktauj}) the limits for the charged
current coupling strength of a neutral heavy lepton with a tau
neutrino.  One can see that, as expected, the tau neutrino sector is
more difficult to probe so the main constraints come from
NOMAD~\cite{Astier:2001ck}, CHARM~\cite{Orloff:2002de}, and
DELPHI~\cite{Abreu:1996pa}.  Regarding the universality constraint it
is obtained from Eq.~(\ref{eq12}) and implies a forbidden horizontal
band in Fig.~\ref{fig:Ktauj}, about 3\% at 90\% C.L.
One should note, however, that the limits for the mixing typically
rely upon extra assumptions on the decay of the neutral heavy lepton,
hence somewhat model-dependent.  
As will be seen below in section~\ref{sec:universality}, the existence
of neutral heavy leptons affects low energy weak decay processes,
where the neutrinos that can be kinematically produced are only the
light ones. From those searches one obtains model--independent
constraints which are also shown in Figs.~(\ref{fig:Kej})
and~(\ref{fig:Kmuj}) as a light horizontal region on the top of the
figures.

\subsection{Weak universality bounds}
\label{sec:universality}

If right--handed neutrinos are the messengers of neutrino mass
generation the corresponding mass eigenstates (also called neutral
heavy leptons) will not be emitted in several weak decays, such as
beta or muon decays, due to kinematics.  Therefore, such decays would
measure different values for the Fermi constant, violating
universality. In particular, for the aforementioned beta and muon
decay, the Fermi constant would be modified to be
\begin{equation}\label{eq1A}
G_{\beta}=G_F\, \sqrt{(NN^\dagger)_{11}}\, 
= G_F\, \sqrt{\alpha_{11}^2}
\end{equation}
and 
\begin{equation}\label{eq1}
G_{\mu}=G_F\, \sqrt{(NN^\dagger)_{11}(NN^\dagger)_{22}}\, 
= G_F\, \sqrt{\alpha_{11}^2(\alpha_{22}^2+|\alpha_{21}|^2)}
\end{equation}
respectively. Since the Fermi coupling constant appears basically in
every weak process, almost any observable should be affected by
non-unitarity, particularly the well measured CKM matrix
elements. Even the pion decay branching ratio
\begin{equation}
R_{\pi} = \frac{\Gamma (\pi^+ \to e^+\nu)}{\Gamma (\pi^+ \to \mu^+\nu)} 
\end{equation}
will also be modified by universality violation:
\begin{equation}\label{eq10}
r_\pi= \frac{R_{\pi}}{R^{SM}_{\pi}} =
\frac{(NN^\dagger)_{11}}{(NN^\dagger)_{22}} =
\frac{\alpha_{11}^2}{\alpha_{22}^2 + |\alpha_{21}|^2}  .
\end{equation} 
Universality constraints derived from the CKM matrix
elements~\cite{Abada:2013aba} as well as from pion
decay~\cite{Abada:2012mc} has been extensevily analyzed and dedicated
studies have been devoted to this
subject~\cite{Gronau:1984ct,Nardi:1994iv,Langacker:1988ur,Antusch:2014woa,Atre:2009rg}. Using
the above prescriptions for the Fermi constants, and based on the
experimental measurements for the CKM matrix
elements~\cite{Agashe:2014kda} and for the pion branching
ratio~\cite{Czapek:1993kc} one obtains, at $90\%$ C.L.,
\begin{eqnarray}
1-\alpha_{11}^2 &<& 0.0130 \nonumber \, , \\
1-\alpha_{22}^2-|\alpha_{21}|^2 &<& 0.0012 \, .
\label{eq:universality-bounds}
\end{eqnarray}
For the case of $\mu-\tau$ universality test, one can use the few
experimental data available~\cite{Aubert:2009qj} in order to get a
constraint
\begin{equation}\label{eq12}
\frac{(NN^\dagger)_{33}}{(NN^\dagger)_{22}}=0.9850 \pm 0.0057 \,.
\end{equation}

\subsection{High energy coliders}
\label{sec:high-energy-coliders}

More interesting is the possibility that right-handed neutrinos are
the messengers of neutrino mass generation. In this case the smallness
of neutrino masses may severely restrict the magnitudes of the
right-handed neutrino admixtures and hence close their potential
signatures at collider experiments like the LHC.
This is expected in the high-scale Type-I seesaw. However, these
limitations can be avoided within a broad class of low-scale seesaw
realizations, such as the inverse
seesaw~\cite{Mohapatra:1986bd,gonzalezgarcia:1989rw,Akhmedov:1995ip,Akhmedov:1995vm,Malinsky:2005bi}
and linear
seesaw~\cite{Boucenna:2014zba,Dev:2009aw,Dev:2013oxa,Drewes:2015iva}
mechanisms.

Within the standard \SM model heavy neutrinos in the TeV range, would
be produced only through relatively small mixing effects, since they
are mainly isosinglets.
Still signatures associated with such heavy heavy neutrinos can
searched for directly at accelerator experiments, like LEP and the
LHC.
Indeed, as can be seen from Figs.~\ref{fig:Kej} and~\ref{fig:Kmuj} the
restrictions from the LHC are weaker than what would be expected from
unitarity violation bounds in this mass range. LHC constraints are
currently absent in the case of the tau-flavor, as seen in
Fig.~\ref{fig:Ktauj}, though in principle one expects that future LHC
runs will improve the situation~\cite{Degrande:2016aje}.

In contrast, the above limitation can be avoided in extended
electroweak models with larger gauge groups. In such case the extra
kinematically accessible gauge bosons provide a production portal for
the heavy neutrinos which may be copiously produced, an can also give
rise to \lfv signatures. As an example one can have left-right
symmetric models, which lead to processes with lepton flavour
violation at high energies~\cite{AguilarSaavedra:2012fu,Das:2012ii}.
Another interesting extension are models with \TrTrOne gauge
symmetry~\cite{Boucenna:2014ela,Boucenna:2014dia,Boucenna:2015raa,Vien:2016tmh},
which have many interesting features as well as experimental
signatures. 

\subsection{Neutrinoless double beta decay}
\label{sec:neutr-double-beta}

\begin{figure}[!htb]
  \centering
\includegraphics{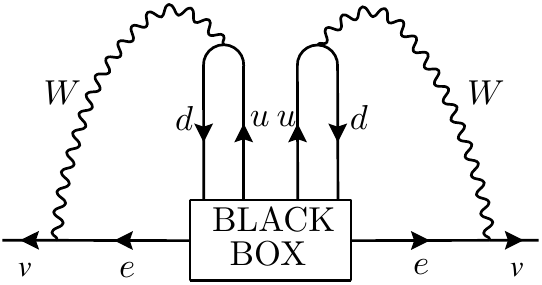}
\caption{Neutrinoless double beta decay implies that neutrinos are own
  anti--particles~\cite{Schechter:1981bd}.}
 \label{fig:bbox}
\end{figure}

Since current knowledge of the neutrino mixing matrix comes
exclusively from oscillation experiments, there is no information on
the absolute neutrino mass scale, neither on the Majorana phases.
A possible detection of neutrinoless double beta decay would imply
that neutrinos are their own anti--particles~\cite{Schechter:1981bd}
irrespective of the underlying origin of neutrino mass, as illustrated
in Fig.~\ref{fig:bbox}.
It would also provide complementary information inaccessible within
oscillation studies and the searches described
above~\cite{Deppisch:2012nb,Rodejohann:2011mu}. The effective Majorana
neutrino mass will be given by~\cite{Deppisch:2012nb}
\begin{equation}
\label{eq:mass-nu00}
\vev{m} = |\sum_j (U^{n\times n}_{ej})^2 m_j | . 
\end{equation}
Notice that j runs only for the light neutrinos. The presence of the
heavy neutrinos modifies the mixing matrix entries as:
$U^{n\times n}_{ej}=\alpha_{11}U_{ej}$.
Moreover, heavy neutrino exchange will lead short-range \znbb
contributions. 
These will be suppressed due to their mixing in the rectangular
charged current mixing matrix, and will involve a different mass
dependence, since the \znbb amplitude is proportional to
\begin{equation}
{\cal A} \propto \frac{m_j}{q^2 - m_j^2}.
\end{equation}
For neutrino masses above the typical neutrino momentum (around
$0.1$~GeV), the amplitude will be inversely proportional to the mass
of the heavy state. We show in Fig.~(\ref{fig:Kej}) the limit for
$S_{ej}$ obtained from $^{76}$~Ge for a single massive isosinglet
neutrino~\cite{Mitra:2011qr,Simkovic:2010ka}. We stress that this
limit holds only if neutrinos have Majorana nature, a restriction that
is not required for the other constraints shown on the same figure.

\section{Future tests of non-unitarity in neutrino 
oscillations}
\label{sec:prob-non-unit}

As discussed above, the presence of extra heavy leptons modifies the
standard form of the leptonic mixing matrix.  For example, leptonic
mixing, as well as CP violation, may take place even in the limit where
neutrinos become strictly massless~\cite{valle:1987gv,branco:1989bn}.
This leads to the possibility of \lfv and CP violating
processes~\cite{bernabeu:1987gr,rius:1989gk}. 
Moreover, it leads to new conceptual possibilities for neutrino
propagation which could have dramatic implications in astrophysical
environments~\cite{valle:1987gv,nunokawa:1996tg,grasso:1998tt}.

Non-standard properties such as unitarity violation in neutrino mixing
could also be probed in laboratory studies.
Indeed the survival and conversion neutrino oscillation probabilities
should be modified to 
\begin{eqnarray}
P_{\alpha \beta} = \sum^3_{i,j} N^*_{\alpha i}N_{\beta i}N_{\alpha j}N^*_{\beta j} &-& 
        4 \sum^3_{j>i} Re\left[
        N^*_{\alpha j}N_{\beta j}N_{\alpha i}N^*_{\beta i}\right] 
        \sin^2\left(\frac{\Delta m^2_{ji}L}{4E}\right)  \nonumber \\
&+& 
        2 \sum^3_{j>i} Im\left[
        N^*_{\alpha j}N_{\beta j}N_{\alpha i}N^*_{\beta i}\right] 
        \sin\left(\frac{\Delta m^2_{ji}L}{2E}\right)  .
\end{eqnarray}

\begin{figure}
\begin{center}
\includegraphics[width=0.45\textwidth]{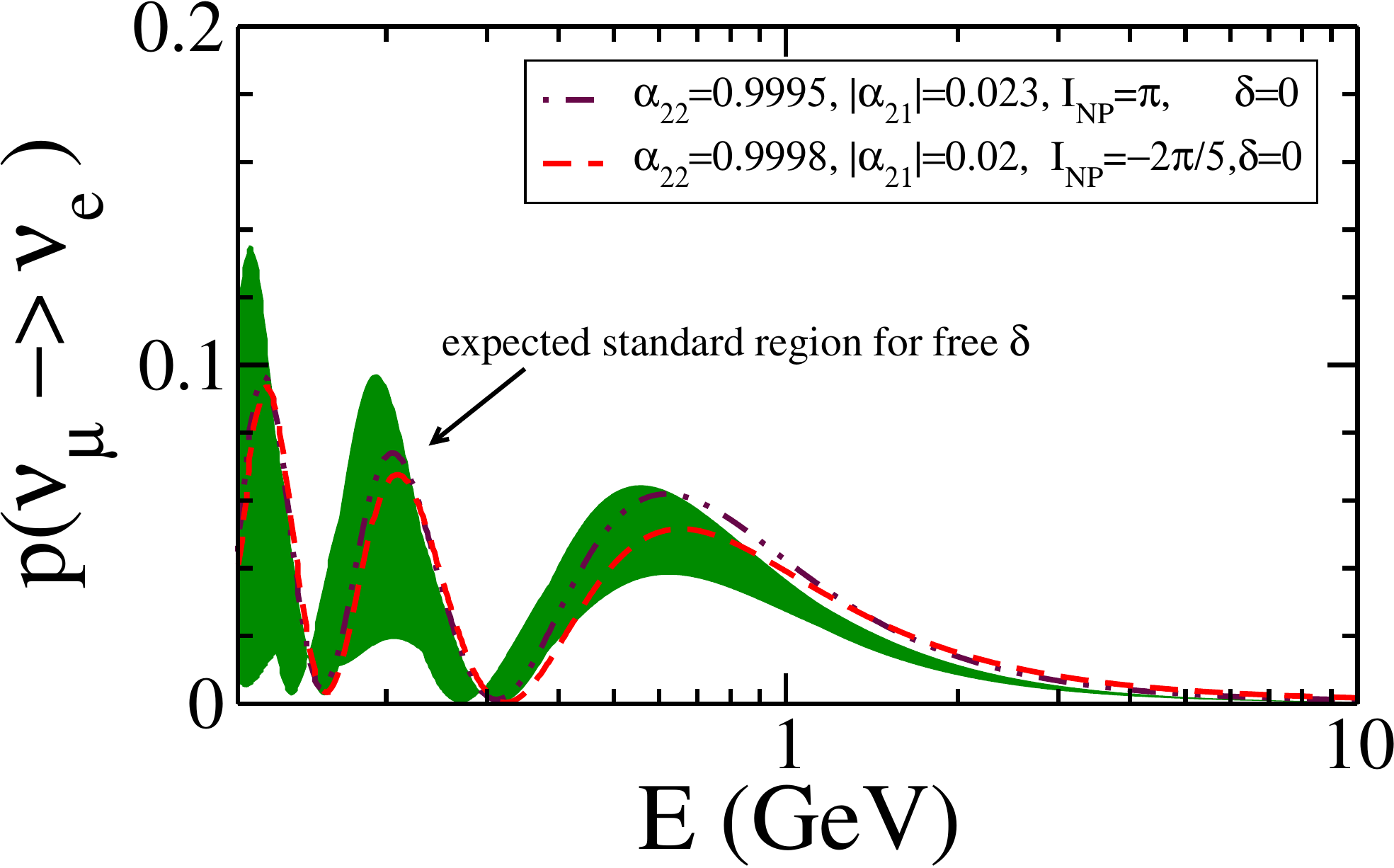}
\caption{Conversion probability for a fixed baseline of 295~km.  The
  green region shows the standard conversion probability with all the
  parameters fixed to the current best fit value, except for the CP
  phase $\delta = -I_{123}$ which is left free. The non-unitary
  case is illustrated with vanishing standard CP phase and two
  different choices of the new physics parameters: as indicated.}
\label{fig:pmue1}
\end{center}
\end{figure}

For instance, for the case of muon to electron neutrino conversion the
transition probability, in vacuum, will be given by the very simple
approximate formula~\cite{Escrihuela:2015wra}
\begin{equation}
P_{\mu e} = \alpha_{11}^2|\alpha_{21} |^2 +
 (\alpha_{11}\alpha_{22})^2 P^{3\times3}_{\mu e}
+  \alpha_{11}^2 \alpha_{22}|\alpha_{21}|  P^{I}_{\mu e} .
\end{equation}
Here, $P^{3\times 3}_{\mu e}$ is the standard three-neutrino
oscillation formula~\cite{Nunokawa:2007qh}. Notice that in this case a
new constant term appears, $\alpha_{22}^2|\alpha_{21}|^2$, that
accounts for the effect of non-unitarity at zero distance, associated
to the effective non-orthogonality of the weak eigenstate
neutrinos~\cite{valle:1987gv}.
Finally, the new term, $P^{I}_{\mu e}$, will depend on two different
phases: the standard CP phase $\delta$ characterizing three-neutrino
oscillations, and an additional CP phase associated to the new
physics, $I_{NP}$, given by the argument of the complex parameter
$\alpha_{21}=|\alpha_{21}|exp(I_{NP})$:
\begin{eqnarray}
P^{I}_{\mu e} & = &
-2 
   \bigg[
   \sin(2\theta_{13}) \sin\theta_{23} 
   \sin\left( \frac{\Delta m^2_{31}L} {4E_\nu}\right)
   \sin\left(\frac{\Delta m^2_{31}L}{4E_\nu} + I_{NP}- I_{123}\right) \bigg]
\nonumber \\ 
  & - &  \cos\theta_{13} \cos\theta_{23} 
  \sin(2\theta_{12}) 
   \sin\left(\frac{\Delta m^2_{21}L}{2E_\nu}\right)
  \sin(I_{NP})
   .
\end{eqnarray}
Therefore, the effect of non-unitarity can be described by four real
parameters : $\alpha_{11}$, $\alpha_{22}$, $|\alpha_{21}|$ plus the
phase $I_{NP}$.

The conversion probability for the neutrino appearance chahnel in the
T2K experiment, characterized by a 295~km
baseline~\cite{Abe:2011sj,Abe:2013hdq}, is given in
Fig.~(\ref{fig:pmue1}). The green region shows the standard conversion
probability with all the oscillation parameters fixed to their current
best fit value, except for the new CP phase which has been left
free. On the other hand the non-unitary case is illustrated with a
standard CP phase equal to zero and two different choices of the new
physics parameters: $\alpha_{11} = 1$, $\alpha_{22}= 0.9995$,
$|\alpha_{21}|=0.023$, and $I_{NP}= \pi$ (maroon dashed-dotted
line); and $\alpha_{11} = 1$, $\alpha_{22}= 0.9998$,
$|\alpha_{21}|=0.02$, and $I_{NP}=-2 \pi/5$ (red dashed line).

\begin{figure}
\begin{center}
\includegraphics[width=0.45\textwidth]{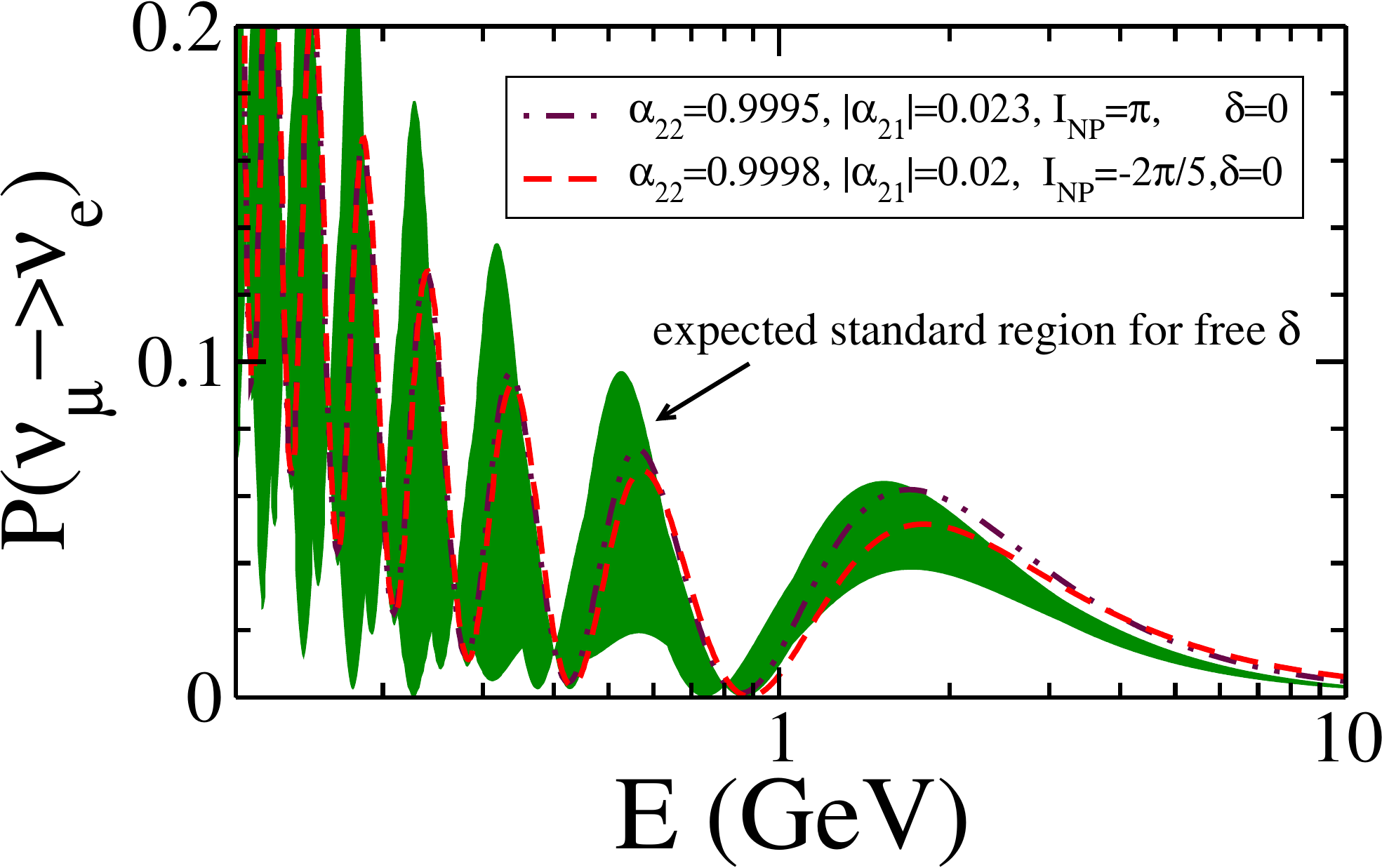}
\caption{Conversion probability for a fixed baseline of 810~km.  The
  green region shows the standard conversion probability with all the
  parameters fixed to the current best fit value, except for the CP
  phase $\delta = -I_{123}$ which is left free. The non-unitary case is
  illustrated with a vanishing standard CP phase and two different
  choices of the new physics parameters, as indicated.}
\label{fig:pmue2}
\end{center}
\end{figure}

Recently the NOvA experiment that has reported a new measurement of
the electron neutrino appearance channel~\cite{Adamson:2016tbq}.
Motivated by this we show, in Fig~(\ref{fig:pmue2}), the behaviour of
the conversion probability for the case of a 810~km baseline.
The green region shows the standard conversion probability with all
the parameters fixed to the current best fit value, except for the new
CP phase $\delta = -I_{123}$, which has been left free. The
non-unitary case is illustrated with a standard CP phase equal to zero
and two different choices of the new physics parameters, indicated in
the plot. 

\begin{figure}
\begin{center}
\includegraphics[width=0.45\textwidth]{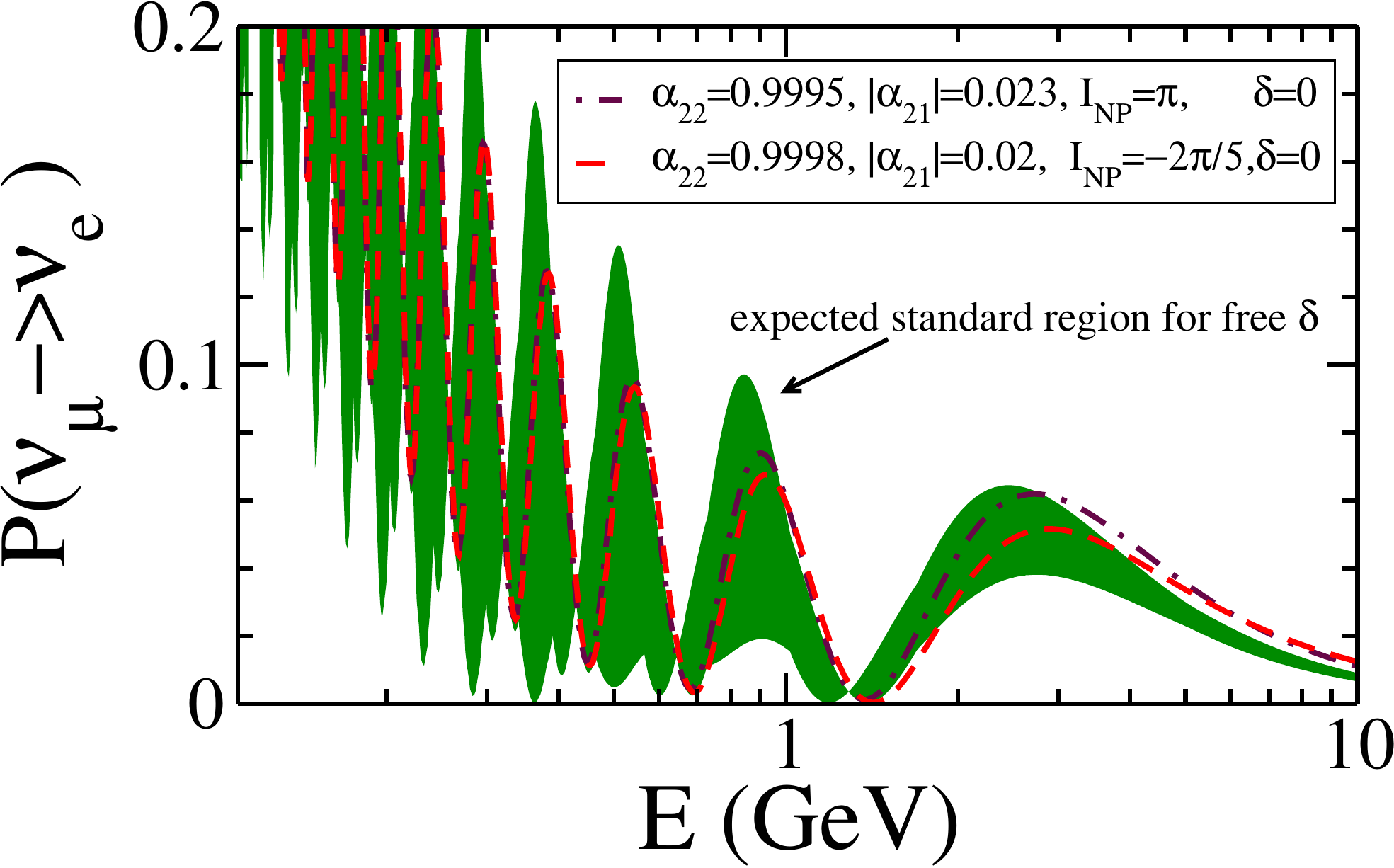}
\caption{Conversion probability for a fixed baseline of 1,300~km.  The
  green region shows the standard conversion probability with all the
  parameters fixed to the current best fit value, except for the CP
  phase $\delta = -I_{123}$ that has been left free. The non-unitary
  case is illustrated with a standard CP phase equal to zero and two
  different choices of the new physics parameters, as indicated.}
\label{fig:pmue3}
\end{center}
\end{figure}

Finally, in Fig.~(\ref{fig:pmue3}) we plot the conversion probability
for a baseline of 1,300~km, relevant for the future DUNE
proposal~\cite{Acciarri:2015uup}. Although we have neglected matter
effects, our results illustrate pretty well the main qualitative
point.
The shaded band in this figure is the standard region for the central
values of the current neutrino oscillation parameters as reported in
Ref.~\cite{Forero:2014bxa}, leaving the CP phase completely free. 
The panels also show two survival probabilities including
non-unitarity effects, for two particular choices of parameters. In
these two new physics cases we have set the standard CP phase to zero
and we have taken a non-zero value for the extra phase $I_{NP}$. 
One sees that, in all the above cases, our results are suggestive of
the fact that there is room for a degeneracies between standard
oscillations and new physics associated to non-unitarity or ``seesaw''
effects. These will make it difficult to extract the standard CP
effects and will certainly be one of the challenges which future
neutrino experiments will have to face.

\section{Conclusions}
\label{Conclusions}

Here we gave a brief summary of the theoretical interpretation of
current neutrino oscillation data within the three--neutrino paradigm.
If neutrinos get their mass \emph{a la seesaw}, we may expect both
direct and/or indirect effects associated to the neutrino mass
generation messengers, e.g. the heavy right-handed neutrinos.
These could be indicative of the simplest next step in particle
physics.
Insofar as oscillations are concerned, we pointed out the case for
unitarity violation in the ``PMNS'' matrix. Experiments are usually
interpreted within the unitary approximation.  However, we illustrated
how CP violation studies could confuse genuine CP violation with
effects associated with unitarity deviations.
Taking up this challenge would shed light on the mass scale of \lnv
and neutrino mass generation within the seesaw mechanism.
Besides opening the stage for new physics, refined neutrino oscilation
studies might also pave the way towards the understanding of at least
some of the current puzzles facing modern cosmology. The reader is
addressed to~\cite{Sierra:2014sta,Boucenna:2014uma,Smoot:2014era} for
examples of possible cosmological implications of neutrino mass
generation.

The interpretation of neutrino data has also been considered in terms
of sub-Fermi strength non-standard interactions of a more generic type
than the non-unitarity effects considered
above~\cite{guzzo:2001mi,Ohlsson:2012kf,Miranda:2015dra,deGouvea}.
Laboratory oscillation studies not only give crucial confirmation of
the oscillation hypothesis, but also establish the robustness of large
mixing angle solar neutrino
oscillations~\cite{Maltoni:2002aw,pakvasa:2003zv}.  One exception is
the existence of a large mixing solution, in the dark side, which
still survives~\cite{miranda:2004nb,Escrihuela:2009up}.  Here we have
focussed on the simplest manifestation of non-standard neutrino
propagation, namely that which comes from the non-unitary form of the
lepton mixing matrix and characterizes Type I seesaw schemes.

\section*{Acknowledgements}
This work is supported by the Spanish grants FPA2014-58183-P,
Multidark CSD2009-00064, SEV-2014-0398 (MINECO), PROMETEOII/2014/084
(Generalitat Valenciana), and the CONACyT grant 166639.


\bibliographystyle{elsarticle-num} 


\end{document}